\documentclass[preprint]{JHEP3} 

\usepackage{amsmath}

\def\beq{\begin{equation}}
\def\eeq{\end{equation}}
\def\bea{\begin{eqnarray}}
\def\eea{\end{eqnarray}}
\def\bec{\begin{center}}
\def\eec{\end{center}}
\def\nn{\nonumber}

\newcommand{\cN}{{\cal N}}
\newcommand{\Tr}{{\rm Tr\;}}
\newcommand{\lambdab}{\overline{\lambda}}

\newcommand{\cA}{{\cal A}}
\newcommand{\cAb}{{\overline{\cal A}}}
\newcommand{\cF}{{\cal F}}
\newcommand{\cFb}{{\overline{\cal F}}}
\newcommand{\cD}{{\cal D}}
\newcommand{\cDb}{{\overline{\cal D}}}
\newcommand{\cP}{{\cal P}}
\newcommand{\cQ}{{\cal Q}}
\newcommand{\cU}{{\cal U}}
\newcommand{\cUb}{{\overline{\cal U}}} 

\newcommand{\bn}{{\bf n}}
\newcommand{\bx}{{\bf x}}
\newcommand{\bk}{{\bf k}}
\newcommand{\bp}{{\bf p}}

\newcommand{\hatbmu}{\widehat{\boldsymbol {\mu}}}
\newcommand{\hatbnu}{\widehat{\boldsymbol {\nu}}}

\newcommand{\etat}{\widetilde{\eta}}
\newcommand{\psit}{\widetilde{\psi}}
\newcommand{\chit}{\widetilde{\chi}}

\newcommand{\hf}{\frac{1}{2}}
\newcommand{\qtr}{\frac{1}{4}}

\title{A Euclidean Lattice Formulation of $D=5$ Maximally Supersymmetric Yang-Mills Theory}

\author{Anosh Joseph \\
Department of Applied Mathematics and Theoretical Physics (DAMTP) \\
University of Cambridge \\
Cambridge, CB3 0WA \\
United Kingdom \\
anosh.joseph@damtp.cam.ac.uk}

\abstract{We construct lattice action for five-dimensional maximally supersymmetric Yang-Mills theory. This supersymmetric lattice formulation can be used to explore the non-perturbative regime of the continuum target theory, which has a known gravitational dual.}


\preprint{DAMTP-2016-32}

\begin{document}

\section{Introduction}
\label{sec:intro}

Maximally supersymmetric Yang-Mills (MSYM) theory in five dimensions has generated a lot of interest in the recent years \cite{Douglas:2010iu, Lambert:2010iw, Witten:2011zz, Bolognesi:2011nh, Tachikawa:2011ch, Kim:2011mv, Lambert:2011eg, Young:2011aa, Kallen:2012cs, Hosomichi:2012ek, Kallen:2012va, Kim:2012ava, Kallen:2012zn, Bak:2012ct, Fukuda:2012jr, Bern:2012di, Kim:2012qf, Lambert:2012qy, Minahan:2013jwa, Bak:2013bba}. This theory takes part in the AdS-CFT correspondence and its finite temperature properties have been investigated recently in the planar limit. Both the gauge theory and supergravity calculations show that the free energy has $N^3$ scaling behaviour in the large-$N$ limit \cite{Klebanov:1996un, Henningson:1998gx}. The five-dimensional gauge theory has a UV completion in the six-dimensional $(2, 0)$ superconformal theory. The six-dimensional theory is also an interesting theory in its own. In fact, it is one of the most intriguing theories in Nahm's classification of superconformal theories \cite{Nahm:1977tg}. It has no description in terms of a Lagrangian but its existence can be shown using arguments based on string theory and M-theory \cite{Witten:1995zh}. The six-dimensional theory, compactified on $S^1$ leads to the $D=5$ MSYM theory. It was conjectured recently \cite{Douglas:2010iu, Lambert:2010iw} that $D=5$ MSYM captures all the degrees of freedom of the parent six-dimensional theory suggesting that these two theories are the same.

We can learn a lot about the six-dimensional $(2,0)$ theory through the AdS-CFT correspondence, where the $(2, 0)$ theory is conjectured to be dual to M-theory (or supergravity) on $AdS_7 \times S^4$ background \cite{Maldacena:1997re}. Supergravity calculations reveal that the free energy of the $(2, 0)$ theory has $N^3$ scaling behaviour \cite{Klebanov:1996un, Henningson:1998gx}. One can compactify the $(2, 0)$ theory on a circle to obtain the $D=5$ MSYM. Since the $N^3$ behaviour remains in the supergravity dual after compactification, one might expect to find some indication of it in the five-dimensional MSYM. In Ref. \cite{Kallen:2012zn} the $N^3$ scaling behaviour has been reproduced in the five-dimensional theory, at large 't Hooft coupling, through calculations based on large-$N$ matrix model, agreeing with the supergravity prediction\footnote{In Ref. \cite{Kallen:2012zn} the authors found a numerical mismatch in the coefficient of the free energy relation, though it scales like $N^3$ as predicted by supergravity. This mismatch is intriguing and it may be related to the way in which the compactification radius $R_6$ and the five-dimensional Yang-Mills coupling $g_5$ are connected. Also, since the five-dimensional theory is not superconformal, there is no unique canonical way to put it on the five-sphere and this may lead to a numerical mismatch for the coefficient.}.

In the five-dimensional theory the square of the gauge coupling has the dimension of length and a simple power counting suggests that the theory is not perturbatively renormalizable. The six-dimensional theory is believed to be UV finite suggesting that the five-dimensional theory is also UV finite in order for the conjecture in Refs. \cite{Douglas:2010iu, Lambert:2010iw} to hold. In Ref. \cite{Bern:2012di} the authors computed the six-loop four-point correlation function of the five-dimensional theory in the planar limit and showed that it is divergent. This indicates that the connection between the five- and six-dimensional theories is not so straightforward and the claim that these two theories are equivalent may not be true\footnote{In Ref. \cite{Papageorgakis:2014dma} the authors provide a mechanism in which the perturbative divergences could be cancelled by soliton contributions.}. Though this conjecture may not be true due to the existence of the UV divergences in $D=5$ MSYM, one can still expect that the supersymmetric observables of these two theories match. For example, one can compute the expectation values of the Wilson loop in the $D=5$ MSYM, and through the relation between the radius of compactification and the Yang-Mills coupling, one could compute the Wilson surface in the $(2, 0)$ theory.

We also note that the six-dimensional $(2,0)$ theory when compactified on a torus gives rise to the famous four-dimensional $\cN=4$ super Yang-Mills. This theory also takes part in the AdS-CFT correspondence. Numerical results showing some evidence for the correspondence was reported recently in Ref. \cite{Honda:2013nfa}. There the authors compute the correlation functions of chiral primary operators through Monte Carlo simulations to test the predictions of the correspondence, in the strong coupling limit. It is encouraging to see that their results are consistent with the predictions from the correspondence, in particular, the violation of the non-renormalization property observed in the four-point function has the same order of magnitude as predicted by the AdS-CFT correspondence. 

The paper is organized as follows. In Sec. \ref{sec:5d_n2_msym} we write down the maximally supersymmetric Yang-Mills theory in five dimensions and its twisted version, which has a lattice compatible form. We provide the lattice construction of the theory in Sec. \ref{subsec:sym-lattice-n2d5}, following the rules of geometric discretization for twisted bosons and fermions. We also reveal several unique characteristics of the lattice formulation of this theory: it preserves one supersymmetry charge exact at finite lattice spacing, it is gauge invariant on the lattice and it avoids the problem of spectrum doubling on the lattice. In Sec. \ref{sec:conclusion} we provide discussion and conclusions. 

\section{Five-dimensional MSYM in the Continuum}	
\label{sec:5d_n2_msym}

The five-dimensional MSYM theory can be obtained by dimensional reduction of $\cN=1$ Yang-Mills theory in ten dimensions. Since we are interested in the lattice formulation of this theory, on a flat Euclidean spacetime lattice, we will work with Euclidean signature.

The action of the ten-dimensional $\cN=1$ Yang-Mills theory with gauge group $G$ has the following well known form 
\beq
\label{eq:10d_action}
S_{10} = \frac{1}{g_{10}^2} \int d^{10} x ~\Tr \left(\qtr F_{IJ} F_{IJ} - i \lambdab \Gamma_I D_I \lambda \right),
\eeq
with $g_{10}$ denoting the gauge coupling, $F_{IJ}$, $I,J = 0, \cdots, 9$, the ten-dimensional curvature associated with the gauge field $A_I$, and $\lambda$ the gaugino. All fields of the theory transform in the adjoint representation of the gauge group. 

We dimensionally reduce the above action to five dimensions by restricting all fields to be independent of the coordinates $x^6, \cdots, x^{10}$. The components of the ten-dimensional gauge field are separated into components of a five-dimensional gauge field and a set of five adjoint scalars
\beq
A_I = (A_m, \phi_j),
\eeq
with $m = 1,\cdots, 5$ and $j = 6, \cdots, 10$.

The five-dimensional theory resulting from dimensional reduction has $\cN=2$ supersymmetry, with a Euclidean rotation group $SO_E(5)$ and an R-symmetry group $SO_R(5)$. The action of the five-dimensional theory is
\bea
S_5 &=& \frac{1}{g_5^2} \int d^5x ~\Tr \Big(\qtr F_{mn} F_{mn} + \hf D_m \phi_j D_m \phi_j + \qtr [\phi_j, \phi_k] [\phi_j, \phi_k] \nn \\
&& \quad \quad \quad \quad \quad \quad \quad \quad \quad -i\lambdab^{aX}(\gamma^m)_a^{~~b} D_m \lambda_{bX} - \lambdab^{aX} (\gamma^j)_X^{~Y} [\phi_j, \lambda_{aY}] \Big),
\eea  
where the spinor indices $(a,b)$ correspond to the Euclidean rotation group and the indices $(X,Y)$ that of the R-symmetry group.

Since the R-symmetry group of the five-dimensional theory is as large as its Euclidean spacetime rotation group we can {\it maximally twist} this theory to obtain a twisted version of the theory. In five dimensions the twisting process is unique and it leads to a twisted theory with the B-model type twist. The topologically twisted version of this theory has been constructed in Ref. \cite{Geyer:2002gy} and we will follow the same procedure for the twisted theory in the continuum. We also note that the twisted version of the five-dimensional theory has gained a lot of interest after Witten's work on Khovanov homology of knots via gauge theory \cite{Witten:2011zz}. In order to obtain the twisted version of this theory we take the new rotation group of the theory to be the diagonal subgroup of $SO_E(5) \times SO_R(5)$. This amounts to identifying the spinor indices, $a$ and $X$, and the spacetime and internal symmetry indices $m$ and $j$. 

The action of the five-dimensional theory takes the following form after twist
\bea
\label{eq:5d_twisted_action}
S_5 &=& \frac{1}{g_5^2} \int d^5x ~\Tr \Big(\qtr \cFb_{mn} \cF_{mn} + \hf D_m \phi_m D_n \phi_n \nn \\
&& \quad \quad \quad \quad \quad \quad \quad -i\lambdab^{AC}(\gamma^m)_A^{~~B} D_m \lambda_{BC} - \lambdab^{CA} (\gamma^m)_A^{~~B} [\phi_m, \lambda_{CB}] \Big),
\eea  
where $\phi_m$ is now promoted to a five-dimensional vector field after twisting. It is natural to combine the two vector fields $A_m$ and $\phi_m$ to form a complexified gauge field
\bea
\cA_m &=& A_m + i\phi_m,~~~~\cAb_m = A_m - i\phi_m.
\eea

The complexified field strengths appearing in the above action are defined as
\bea
\cF_{mn} &\equiv& [\cD_m, \cD_n],~~~~\cFb_{mn} \equiv [\cDb_m, \cDb_n],
\eea
with the complexified covariant derivatives 
\bea
\cD_m &\equiv& \partial_m + [\cA_m, \cdot~],~~~~\cDb_m \equiv \partial_m + [\cAb_m, \cdot~].
\eea

In the twisted theory it is more appropriate to write down the spinor fields using tensor notation since the degrees of freedom with half-integer spin are now changed to the ones with integer spin (Grassmann-odd p-forms) under the new rotation group. The twisted spinor field can be decomposed in the following way \cite{Geyer:2002gy}
\bea
\label{eq:decompo}
\lambda_{AB} = \frac{1}{2 \sqrt{2}} \Big(\hf (\sigma^{mn})_{AB} \chi_{mn} + (\gamma^m)_{AB} \psi_m - \epsilon_{AB} \eta \Big),~~\lambda^{AB} = \epsilon^{AC} \epsilon^{BD} \lambda_{CD},
\eea
where the fermion fields $\eta, \psi_m$ and $\chi_{mn}$ are scalar, vector and anti-symmetric tensor fields respectively, and they uniquely capture the sixteen degrees of freedom of the original ten-dimensional spinor field. Here $(\sigma_{mn})_{AB}$ denote the ten generators of the $Sp(4) \sim SO(5)$ rotations, $(\gamma^m)_{AB}$ are hermitian $SO(5)$ matrices and $\epsilon^{AB}$ is a real and antisymmetric invariant tensor of $Sp(4)$. 

Inserting the spinor decomposition given in Eq. \eqref{eq:decompo} into the action Eq. \eqref{eq:5d_twisted_action} and introducing a bosonic auxiliary field $d$, we arrive at the following form of the five-dimensional twisted action
\bea
S_5 &=& \frac{1}{g_5^2} \int d^5x ~\Tr \Big(\qtr \cFb_{mn} \cF_{mn} - d D_m \phi_m - \hf d^2 - i \chi_{mn} \cD_m \psi_n - i \psi_m \cDb_m \eta \nn \\
&& \quad \quad \quad \quad \quad \quad \quad \quad \quad \quad - \frac{i}{8} \epsilon_{mncde} \chi_{de} \cDb_c \chi_{mn} \Big).
\eea  

Integrating out the auxiliary field $d$ using its equation of motion
\beq
d = -D_m \phi_m = \frac{i}{2}[\cD_m, \cDb_m],
\eeq
the twisted action takes the form
\bea
\label{eq:twisted-5d-action}
S_5 &=& \frac{1}{g_5^2} \int d^5x ~\Tr \Big(\qtr \cFb_{mn} \cF_{mn} - \frac{1}{8} [\cD_m, \cDb_m]^2 - i \chi_{mn} \cD_m \psi_n - i \psi_m \cDb_m \eta \nn \\
&& \quad \quad \quad \quad \quad \quad \quad \quad \quad \quad - \frac{i}{8} \epsilon_{mncde} \chi_{de} \cDb_c \chi_{mn} \Big).
\eea  

The supersymmetry charges of the original theory also undergo a decomposition similar to that of the fermions given in Eq. \eqref{eq:decompo}, resulting in scalar, vector and anti-symmetric tensor supercharges, $\cQ$, $\cQ_m$ and $\cQ_{mn}$, respectively. The supersymmetry algebra can be rewritten in terms of the twisted supercharges. The twisted supersymmetry algebra contains a subalgebra in which the scalar supercharge is strictly nilpotent: $\cQ^2 = 0$. Since this subalgebra does not generate any translations we can easily transport the theory on to the lattice by preserving one supersymmetry charge exact at finite lattice spacing.

The twisted action can be written as a sum of $\cQ$-exact and $\cQ$-closed terms
\beq
\label{eq:5d_n2_action}
S_5 = \cQ \Lambda - \frac{1}{g_5^2} \int d^5x ~\Tr \frac{i}{8} \epsilon_{mncde} \chi_{de} \cDb_c \chi_{mn},
\eeq
where $\Lambda$ is a functional of the fields
\beq
\Lambda = \frac{1}{g_5^2}  \int d^5x ~\Tr \Big(\frac{i}{4} \chi_{mn} \cF_{mn} - \eta D_m \phi_n - \hf \eta d \Big).
\eeq

It is easy to see that the twisted action is $\cQ$-invariant. The $\cQ$-exact piece vanishes trivially due to the nilpotent property of the scalar supercharge and the vanishing of $\cQ$-closed piece can be shown through the Bianchi identity for covariant derivative
\beq
\epsilon_{mncde} \cDb_c\cFb_{de} = \epsilon_{mncde} [\cDb_c,[\cDb_d, \cDb_e]] = 0.
\eeq  

The supersymmetry transformations generated by the scalar supercharge have the following form
\bea
\cQ \cA_m &=& \psi_m, \\
\cQ \psi_m &=& 0, \\
\cQ \cAb_m &=& 0, \\
\cQ \chi_{mn} &=& -i \cFb_{mn}, \\
\cQ \eta &=& d, \\
\cQ d &=& 0,
\eea
and it is easily seen that the $\cQ$ supersymmetry charge is strictly nilpotent on the twisted fields.

The five-dimensional twisted MSYM theory discussed above is a higher dimensional analogue of the class of models introduced by Blau and Thompson \cite{Blau:1996bx} and in fact the action Eq. \eqref{eq:5d_n2_action} has a strong resemblance to the topologically twisted action of the three-dimensional $\cN=4$ SYM theory. After dimensional reduction to four dimensions one obtains the action of the B-twisted model, which is nothing but the twisted version of $\cN=4$ SYM theory originally constructed by Marcus \cite{Marcus:1995mq}, which takes part in the geometric Langlands program \cite{Kapustin:2006pk}. From the fixed points of the fermionic supersymmetry transformations given above we see that the action of the five-dimensional theory localizes on to the moduli space of complexified flat connections. 
  
The twisted action given in Eq. \eqref{eq:5d_n2_action} can be dimensionally reduced to four dimensions by splitting the coordinates into $x_m = (x_\mu, x_5)$, with $\mu = 1, \cdots, 4$, and restricting all fields to be independent of $x_5$. After renaming the fifth component of the fields the following way
\beq
\cAb_5 = B,~~\cA_5 = \bar{B},~~\chi_{\mu 5} = \psit_\mu,~~\psi_5 = \etat,
\eeq 
and retaining the other components of the fields as they are we arrive at the twisted action of the $\cN=4$ SYM constructed by Marcus \cite{Marcus:1995mq}

\bea
\label{eq:Marcus_4d}
S_4 &=& \frac{1}{g_4^2} \int d^4x ~\Tr \Big( \qtr \cFb_{\mu \nu} \cF_{\mu \nu} + \qtr \cD_\mu B \cDb_\mu \bar{B} + \qtr \cDb_\mu B \cD_\mu \bar{B} - \frac{1}{8} [B, \bar{B}]^2 \nn \\
&&\quad \quad \quad \quad - d D_\mu \phi_\mu - \hf d^2 - i \chi_{\mu \nu} \cD_\mu \psi_\nu - i \chit_{\mu \nu} \cDb_\mu \psit_\nu + \frac{i}{4} B \{\chi_{\mu \nu}, \chit_{\mu \nu} \} \nn \\
&&\quad \quad \quad \quad - i \psit_\mu \cD_\mu \eta - i \psi_\mu \cDb_\mu \etat - i \bar{B} \{ \psi_\mu, \psit_\mu \} + i B \{\eta, \etat\} \Big),
\eea
where $\chit_{\mu \nu} = \hf \epsilon_{\mu \nu \rho \sigma} \chi_{\rho \sigma}$ is the dual of $\chi_{\mu \nu}$. This action also localizes onto the moduli space of complexified flat connections.

We also note that the four-dimensional action Eq. \eqref{eq:Marcus_4d} is invariant under a $Z_2$ symmetry \cite{Geyer:2002gy}
\bea
\label{eq:Z2-symm-1}
\left( \begin{array}{ccccc}
A_\mu & \phi_\mu & B & \bar{B} & d \end{array} \right) &\to& \left( \begin{array}{ccccc}
A_\mu & -\phi_\mu & B & \bar{B} & -d \end{array} \right), \\
\label{eq:Z2-symm-2}
\left( \begin{array}{cccccc}
\eta & \etat & \psi_\mu & \psit_\mu & \chi_{\mu \nu} & \chit_{\mu \nu} \end{array} \right) &\to& \left( \begin{array}{cccccc}
\etat & \eta & \psit_\mu & \psi_\mu & \chit_{\mu \nu} & \chi_{\mu \nu} \end{array} \right),
\eea
which also maps the supersymmetry charge $Q$ into $\widetilde{Q}$. Thus the four-dimensional theory has two nilpotent supercharges. However, only one supercharge is compatible with the lattice regularization of this theory - see Refs. \cite{Unsal:2006qp, Catterall:2007kn}. We also note that the first ever manifestly supersymmetric lattice regularization of $\cN=4$ MSYM, preserving one strictly nilpotent supercharge, was formulated by Kaplan and \"Unsal in Ref. \cite{Kaplan:2005ta}, using the method of orbifold projection of a supersymmetric matrix model. (See also Refs. \cite{Catterall:2009it, Joseph:2011xy} for reviews.) The $Z_2$ symmetry, Eqs. (\ref{eq:Z2-symm-1}) - (\ref{eq:Z2-symm-2}), does not exist in the five-dimensional $\cN=2$ theory constructed above and thus it has only one nilpotent supersymmetry charge. We will be content with it since this supersymmetry is compatible with the lattice regularization of the theory.

\section{Lattice Construction of 5D MSYM}
\label{subsec:sym-lattice-n2d5}

It is straightforward to discretize the five-dimensional MSYM theory once the action is written in the twisted form given in Eq. (\ref{eq:twisted-5d-action}). One has to address several technicalities once the theory is formulated on the lattice. They include supersymmetry invariance of the lattice action, a gauge invariant construction of the lattice theory, absence of spectrum doublers, taming of possible dangerous counter-terms that can be generated radiatively on the lattice and a consistent continuum limit of the regularized theory. 

We consider a hypercubic lattice made out of unit cells containing five mutually orthogonal basis vectors $\hatbmu_m$ in the positive $x_m$ directions
\bea
\hatbmu_1 &=& (1, 0, 0, 0, 0), \nn \\
\hatbmu_2 &=& (0, 1, 0, 0, 0), \nn \\
\hatbmu_3 &=& (0, 0, 1, 0, 0), \\
\hatbmu_4 &=& (0, 0, 0, 1, 0), \nn \\
\hatbmu_5 &=& (0, 0, 0, 0, 1). \nn 
\eea

In the bosonic sector, this theory has only a complexified vector field with five components. The complex continuum gauge fields $\cA_m$, $m=1, \cdots, 5$, are mapped to complexified Wilson gauge links $\cU_m(\bn)$ and they are placed on the links connecting lattice site $\bn$ to site $\bn + \hatbmu_m$ of the hypercubic lattice. The field $\cUb_m(\bn)$ is placed on the oppositely oriented link, that is, from site $\bn + \hatbmu_m$ to site $\bn$. The complexified field strength is defined in the following way on the lattice
\beq
\cF_{mn}(\bn) \equiv \cD_m^{(+)}\cU_n(\bn) = \cU_m(\bn)\cU_n(\bn + \hatbmu_m) - \cU_n(\bn)\cU_m(\bn + \hatbmu_n).
\eeq
This form of the field strength is antisymmetric in the indices by definition and it reduces to the continuum (complex) field strength in the naive continuum limit. We map the complexified covariant derivative $\cD_m$ into a forward or backward lattice covariant difference operator, $\cD^{(+)}_m$ or $\cD^{(-)}_m$, according to the purpose it serves: it becomes $\cD^{(+)}_m$ if the operation on a field is curl-like and $\cD^{(-)}_m$ if the operation is divergence-like. The discretization rules appropriate for twisted supersymmetric gauge theories are derived in Refs. \cite{Damgaard:2007be, Catterall:2007kn, Damgaard:2008pa}.

The Grassmann-odd fields of the twisted theory have the interpretation as geometric fermions \cite{Rabin:1981qj, Becher:1982ud, Banks:1982iq}. Since the fermions of the twisted theory are p-forms (p $= 0,1,2$), it is natural to place each of them on the p-cell of the five-dimensional hypercubic lattice. The 0-form field $\eta$ is placed at the site $\bn$. The components of the 1-form field $\psi_m(\bn)$ are placed along the links connecting site $\bn$ to site $\bn + \hatbmu_m$. We note that such a prescription makes sense since $\psi_m(\bn)$ is the superpartner of the gauge field $\cA_m(\bn)$. The components of the 2-form fermion $\chi_{mn}(\bn)$ live on the links connecting site $\bn + \hatbmu_m + \hatbmu_n$ to site $\bn$.

The nilpotent scalar supersymmetry charge acts on the lattice fields in the following way
\bea
\cQ \cU_m(\bn) &=& \psi_m(\bn), \\
\cQ \psi_m(\bn) &=& 0, \\
\cQ \cUb_m(\bn) &=& 0, \\
\cQ \chi_{mn}(\bn) &=& -i \Big(\cDb_m^{(+)}\cUb_n(\bn)\Big) = -i \cFb_{mn}^L, \\
\cQ \eta(\bn) &=& d(\bn), \\
\cQ d(\bn) &=& 0.
\eea

We see that the $\cQ$ supersymmetry transforms a bosonic field of one type (site or link) into a fermionic field of the same type (site or link) at the same place on the lattice. 

We need to ensure that the placements of the fields on the hypercubic lattice results in a gauge invariant lattice action. The mappings and orientations of the lattice variables can be easily summarized by providing their gauge transformation properties on the lattice. For $g(\bn)$, a unitary matrix at lattice site $\bn$, which is an element of the gauge group $G$, we have the gauge transformations on the lattice fields
\bea
\label{eq:gauge-t}
\cU_m(\bn) &\rightarrow& g(\bn) \cU_m(\bn) g^{\dagger}(\bn + \hatbmu_m), \nn \\ 
\cUb_m (\bn) &\rightarrow& g(\bn + \hatbmu_m) \cUb_m(\bn) g^{\dagger}(\bn), \nn \\
\eta(\bn) &\rightarrow& g(\bn) \eta(\bn) g^{\dagger}(\bn), \\ 
\psi_m(\bn) &\rightarrow& g(\bn) \psi_m(\bn) g^{\dagger}(\bn + \hatbmu_m), \nn \\ 
\chi_{mn} (\bn) &\rightarrow& g(\bn + \hatbmu_m + \hatbmu_n) \chi_{mn}(\bn) g^{\dagger}(\bn). \nn
\eea
It is clear from above that the gauge transformations depend on the geometric nature of the lattice fields we are considering.   

The covariant forward and backward difference operators act on the lattice fields the following way \cite{Damgaard:2007be, Catterall:2007kn, Damgaard:2008pa}
\bea
\label{eq:discretization-1}
\cD_m^{(+)} f(\bn) &=& \cU_m(\bn) f(\bn + \hatbmu_m) - f(\bn) \cU_m(\bn), \\
\cD_m^{(+)} f_n(\bn) &=& \cU_m(\bn) f_n(\bn + \hatbmu_m) - f_n(\bn) \cU_m(\bn + \hatbmu_n), \\
\cDb_m^{(-)} f_m(\bn) &=& f_m(\bn)\cUb_m(\bn) - \cUb_m(\bn - \hatbmu_m) f_m(\bn - \hatbmu_m), \\
\label{eq:discretization-2}
\cDb_c^{(+)} f_{mn} (\bn) &=& f_{mn}(\bn + \hatbmu_c) \cUb_c(\bn) - \cUb_c(\bn + \hatbmu_m + \hatbmu_n) f_{mn}(\bn).
\eea

It is now straightforward to write down the lattice action of the five-dimensional MSYM. After integrating out the auxiliary field $d$ using its equation of motion
\beq
d(\bn) = -\frac{i}{2} \sum_m \cDb^{(-)}_m \cU_m(\bn),
\eeq
the $\cQ$-exact piece of the action takes the following form
\bea
S_{\cQ-\textrm{exact}} &=& \beta \sum_{\bn, m, n} \Tr \Big(- \qtr \cFb_{mn}^L(\bn) \cF_{mn}(\bn) - \frac{1}{8} \Big(\cDb_m^{(-)}\cU_m(\bn)\Big)^2 \nn \\
&&\quad \quad \quad \quad \quad \quad \quad \quad - i\chi_{mn}(\bn) \cD^{(+)}_m\psi_n(\bn) - i\eta(\bn) \cDb^{(-)}_m\psi_m(\bn) \Big),
\eea
with $\beta$ denoting the lattice coupling. In terms of the `t Hooft coupling $\lambda$ and lattice spacing $a$ it is
\beq
\beta = \frac{N}{\lambda},~~\lambda = \frac{N g_5^2}{a}.
\eeq

Following the set of prescriptions given in Eq. (\ref{eq:gauge-t}) we see that the $\cQ$-exact piece of the action is gauge invariant; each term forms a closed loop on the lattice. The $\cQ$-closed term needs special consideration. It must be modified on the lattice in order to maintain gauge invariance. We note that the covariant difference operator acts on $\chi_{mn}$ field at lattice site $\bn$ the following way
\bea
\cDb^{(+)}_c \chi_{mn} (\bn) &=& \chi_{mn}(\bn + \hatbmu_c) \cUb_c(\bn) - \cUb_c(\bn + \hatbmu_m + \hatbmu_n) \chi_{mn}(\bn).
\eea
The above expression contains open loops connecting lattice sites $\bn + \hatbmu_m + \hatbmu_n + \hatbmu_c$ and $\bn$ in the negatively oriented direction. In order to contract it with $\epsilon_{mncde} \chi_{de}(\bn)$ and form a closed loop on the lattice we introduce the ordered product of link variables along a path connecting lattice sites $\bn$ and $\bn + \hatbmu_m + \hatbmu_n + \hatbmu_c + \hatbmu_d + \hatbmu_e$. Let $C_L$ be a path on the lattice connecting these two sites. Then the path ordered link (POL) is defined as
\beq
\cP_{\rm POL} \equiv \prod_{l \in C_L} \cUb_l, 
\eeq
with $\cUb_l$ denoting a link variable on $C_L$. We choose $\cUb$-fields to form the path ordered link since it is trivially annihilated by $\cQ$-supersymmetry. If one chooses to form the path ordered link using $\cU$-fields the term becomes gauge invariant but it breaks the most important property of the lattice theory, that is, $\cQ$-symmetry at the lattice level.

The $\cQ$-closed term on the lattice now takes the following gauge invariant form
\bea
S_{\cQ-\rm closed} &=& - \frac{i\beta}{8} \sum_{\bn, m,n,c,d,e} \Tr \epsilon_{mncde} \cP_{\rm POL} \chi_{de}(\bn + \hatbmu_m + \hatbmu_n + \hatbmu_c) \cDb^{(+)}_c \chi_{mn}(\bn).~~~~~~~~
\eea

We also note that the above $\cQ$-closed term reduces to its continuum counterpart in the naive continuum limit in which the gauge links are set to unity. 

The lattice action constructed here is invariant under $\cQ$-supersymmetry. The $\cQ$-variation of the $\cQ$-exact term vanishes due to the property $\cQ^2 = 0$. The $\cQ$-closed term vanishes due to Bianchi identity for covariant derivatives on the lattice
\beq
\epsilon_{decmn} \cDb^{(+)}_c \cFb_{mn}^L = 0,
\eeq
with $d, e = 1, \cdots, 5$. (See Ref. \cite{Aratyn:1984bd} for the result derived for the four-dimensional case.)

It should be noted that the fermion fields of the twisted theory do not fill all p-cells of the five-dimensional unit cell, unlike staggered fermions in $D$ spacetime dimensions. In the case at hand, the fermions of the theory fill only half the unit cell. This may raise the question that the theory may have spectrum doublers. In Appendix \ref{sec:absence-doublers} we look at the bosonic and fermionic propagators of the lattice theory and show that the theory does not suffer from spectrum doublers if we follow the discretization prescription given in Eqs. (\ref{eq:discretization-1}) - (\ref{eq:discretization-2}). 

We note that our choice of the path ordered link makes the $\cQ$-closed piece and thus the lattice action non-local. This is the price we have to pay in this particular discretization scheme of the theory in order to preserve gauge invariance and $\cQ$-supersymmetry, and to avoid spectrum doublers at the lattice level. There may also arise a concern about the refection positivity of the lattice action\footnote{We thank the referee for pointing this out.}. One has to check whether the lattice action constructed here satisfies reflection positivity. It is a sufficient, but not necessary, condition for a Hermitian continuum theory. The question on reflection positivity property of twisted supersymmetric lattice Yang-Mills theories in general still remains to be investigated.

We also note that the orbifold projection method of Kaplan and \"Unsal \cite{Kaplan:2005ta} to construct MSYM theories in various dimensions, from a supersymmetric matrix model, cannot give rise to the theory we constructed here. There, in the steps of specifying the desired orbifold projection, the choice of how to embed the $Z_N^D$ symmetry into $SO(10)$ is guided by three principles \cite{Kaplan:2005ta}. Although such a recipe gives the maximum possible dimension for the orbifold projected lattice theory to be $D=5$, the requirement that the daughter theory be local constrains the maximum dimension to be $D=4$. The components of the ${\bf r}$ charges of the orbifold theory assume values $0$ or $\pm 1$ for any of the bosonic or fermionic variables of the mother theory, leading to the requirement that the orbifold projection be chosen such that there are only nearest neighbour interactions on the resulting lattice. With such a definition of the ${\bf r}$ charges one can construct $D$-dimensional local lattices with $2^{4-D}$ unbroken supercharges, which excludes the orbifold construction of the $D=5$ MSYM theory. This particular constraint on the orbifold lattice theories, once relaxed, may have resulted in the modified form of the $\cQ$-closed term in the twisted version of the $D=5$ lattice action we constructed here.

We also note that one may be able to construct a supersymmetric lattice action of the $D=5$ MSYM using the regularization prescription given by Sugino \cite{Sugino:2004uv}. It is also interesting to explore the possibility of a different supersymmetric lattice formulation of this theory from a three-dimensional sixteen supercharge Yang-Mill theory with a fuzzy $S^2$ background, similar to the construction given in Ref. \cite{Hanada:2010kt}. 

\section{Fine-tuning and Restoration of Supersymmetries}
\label{subsec:fine-tuning}

We need to determine the number of parameters that must be fine-tuned in order to recover the desired long-distance effective theory, which gives rise to the $\cN=2$ SYM in the classical continuum limit. We would like to know the possible operators that could be induced on the lattice. Imposing $\cQ$-symmetry and gauge invariance on the lattice could reduce their number drastically. Following the analysis given in Refs. \cite{Catterall:2013roa, Catterall:2014mha} we can write down the most general long distance effective action of the $\cN=2$ theory
\bea
\label{eq:5d_n2_action_gen_coeff}
S_5 &=& \beta \sum_{\bn,m,n,c,d,e} \cQ ~\Tr \Big(\frac{i}{4} \alpha_1 \chi_{mn} \cF_{mn} + \frac{i}{2} \alpha_2 \eta [\cD_m, \cDb_m] - \hf \alpha_3 \eta d \Big) \nn \\
&&\quad \quad - \Tr \frac{i}{8} \alpha_4 \epsilon_{mncde} \chi_{de} \cDb_c \chi_{mn} + \gamma \cQ \left(\Tr \eta \cU_m \cUb_m - \frac{1}{N} \Tr \eta ~\Tr \cU_m \cUb_m \right),
\eea
where $\alpha_i$ with $i = 1, \cdots, 4$ and $\gamma$ are dimensionless numbers taking values $(1, 1, 1, 1, 0)$ in the classical theory.

Acting with $\cQ$ and using the freedom to rescale the fields \cite{Catterall:2014mha} we obtain
\bea
S_5 &=& \beta \sum_{\bn,m,n,c,d,e} \Tr \Big(\qtr \alpha_1 \cFb_{mn} \cF_{mn} + \frac{i}{2} \alpha_1 d [\cD_m, \cDb_m] - \hf \alpha_3'd^2 - i \alpha_1 \chi_{mn} \cD_m \psi_n \nn \\
&&\quad - i \alpha_1 \psi_m \cDb_m \eta - \frac{i}{8} \alpha_1 \epsilon_{mncde} \chi_{de} \cDb_c \chi_{mn} \Big) + \gamma' \Big(\Tr \left( d \cU_m \cUb_m \right) - \Tr \left( \eta \psi_m \cUb_m \right) \nn \\
&&\quad - \frac{1}{N} \Tr d ~\Tr \left( \cU_m \cUb_m \right) + \frac{1}{N} \Tr \eta ~\Tr \left(\psi_m \cUb_m \right) \Big),
\eea  
where
\beq
\alpha_3' = \alpha_3 \left(\frac{\alpha_1}{\alpha_2}\right)^2,~~~\gamma' = \gamma \frac{\alpha_1}{\alpha_2}.
\eeq

Thus we see that a total of at most three fine-tunings will be required: $\alpha'_3 \to \alpha_1$, $\beta$ and $\gamma' \to 0$. Note that the over all coefficient $\alpha_1$ can be absorbed into the lattice coupling $\beta$. 

We could also make a comment about the restoration of R symmetries as we approach the continuum limit. The discrete R symmetries provide a powerful constraint on the relations between $\alpha_i$'s since the operators appear through radiative corrections to the action must respect these additional symmetries. In the four-dimensional $\cN = 4$ Yang-Mills case it was found that the restoration of even a discrete version of the R symmetry, denoted by $R_m$ and $R_{mn}$, is sufficient to guarantee the correct continuum limit. It has the effect of setting $\gamma' \equiv 0$ and the coefficients $\alpha'_3$ and $\alpha_1$ equal to each other in the $\cN = 2$ theory. Thus invariance under the single scalar supercharge $\cQ$ together with invariance under any one of the 15 $R_m$ or $R_{mn}$ symmetries implies that the theory is invariant under {\it all} additional twisted supersymmetries.

The lattice action constructed here does not respect the discrete R symmetries $R_m$ and $R_{mn}$. Thus we cannot draw the same implication as in the continuum, that the coefficients $\alpha_i$ that were discussed above will all be equal. However, if any of the 15 discrete R symmetries emerge in the long distance theory, this equality is sufficient to yield the full $\cN = 2$ supersymmetry at low energies. Thus apart from a renormalization of the overall coefficient $\beta$, no fine-tuning of the lattice action would in this case be required. 

There is some reason to hope that the discrete R symmetries do in fact emerge at low energies. The reason is that the twisting process has combined the $SO_E(5)$ spacetime symmetry with an $SO_R(5)$ symmetry. If the $SO_E(5)$ rotational symmetry of the continuum emerges at low energies, then we also expect to obtain the $SO_R(5)$ symmetry, since they are basically on the same footing as far as twisting is concerned. In that case some of the discrete R symmetries would also be emergent, which is sufficient to guarantee the equality of $\alpha_i$ coefficients.

Thus it is crucial to check the restoration of $SO_E(5)$ invariance and $SO_R(5)$ symmetry as we approach the continuum limit. These properties can be checked through the measurement of correlation functions, which should be related to each other through the symmetries.

We also note that the five-dimensional lattice theory constructed here can exhibit flat directions, a general feature of theories with extended supersymmetry, and they give rise to instabilities in lattice simulations. A way to control them in the simulations is to introduce suitable mass terms to the scalar fields {\it by hand} and then appropriately tune the mass parameters. 

One could add a bosonic mass term of the form
\beq
S_M = \beta \mu^2 \sum_{\bn,m} \Big(\frac{1}{N}\Tr \left(\cU_m(\bn) \cUb_m(\bn) - {\bf I}\right)\Big)^2
\eeq
to the lattice action and simulate the theory at various values of the mass parameter $\mu$. This term is gauge invariant on the lattice but it softly beaks $\cQ$-supersymmetry.

It is also important to make sure that the lattice theory does not suffer from a sign problem associated with the fermion determinant. In the recent work it has been shown that the phase of the Pfaffian is close to zero in MSYM theories in lower dimensions \cite{Catterall:2011aa, Mehta:2011ud, Galvez:2012sv, Catterall:2014vka}.

\section{Discussion and Conclusions}
\label{sec:conclusion}

We have constructed a supersymmetric lattice action of the five-dimensional sixteen supercharge Yang-Mills theory using the methods of topological twisting and geometric discretization. The lattice theory preserves one supersymmetry charge exact at finite lattice spacing. The covariant derivatives of the continuum theory are mapped to forward and backward lattice covariant difference operators through a well defined prescription. The lattice theory is supersymmetric, gauge invariant and free from spectrum doublers. 

The lattice formulation proposed here can be used to explore the non-perturbative regime of MSYM theories in five dimensions at finite gauge coupling and number of colours, and also its parent theory, $(2, 0)$ superconformal theory in six dimensions. It would be interesting to find a nontrivial UV fixed point from the lattice theory for 5D MSYM since the fixed point can give a UV completion and non-perturbative definition of the theory.   

It is important to understand how the lattice theory is renormalized. The continuum theory is known to be perturbatively non-renormalizable and it may find its UV completion non-perturbatively. The new degrees of freedom that are required to define the UV completion of the six-dimensional theory may be already present in the non-perturbative physics of the $D=5$ theory as suggested by Douglas \cite{Douglas:2010iu}. Having a first-principles non-perturbative formulation of this theory would provide us with much more information about these interesting theories. 

The lattice construction formulated here preserves only one supersymmetry charge exact at finite lattice spacing. It is also crucial to understand how the remaining broken supersymmetries are restored as we approach the continuum limit of the theory. Measuring appropriate correlation functions, which should be related to each other through the symmetries, on lattice would provide insight into the restoration of all the supersymmetries as we approach the continuum limit. 

The theory described here takes part in the AdS-CFT correspondence. It is the low energy theory living on a stack of D4 branes. The gravitational dual of this theory is known; it is the supergravity on the near horizon geometry of D4 branes.

The only well established way to find expectation values of observables in $(2, 0)$ superconformal theories is through the AdS-CFT correspondence. Lattice construction of its five-dimensional cousin provides us a tool for non-perturbative investigation of this theory. One of the most important results obtained for $(2, 0)$ theories from their supergravity duals is the $N^3$ scaling behaviour of their free energy and conformal anomaly \cite{Klebanov:1996un, Henningson:1998gx}. The $N^3$ scaling behaviour of the free energy in $D=5$ MSYM theory has been reproduced in Ref. \cite{Kallen:2012zn} in the planar limit. They have also calculated the expectation value of the circular Wilson loop in $D=5$ MSYM. It would be interesting if one could verify this claim through direct lattice simulations of the five-dimensional theory. 

Another interesting supersymmetric observable to compute is the 't Hooft operator, which is the magnetic dual of the Wilson loop operator. The 't Hooft operators in five dimensions are some surface operators corresponding to the world-surface of monopole strings. It could be possible to supersymmetrize and localize them similar to the 't Hooft loop in four dimensions \cite{Giombi:2009ek, Gomis:2011pf}. In the $(2, 0)$ theory there is no difference between 't Hooft and Wilson operators since the theory is self-dual and both are surface operators. Therefore the $D=5$ 't Hooft loop should correspond to a $D=6$ Wilson surface which does not wrap the compactified direction. It would be interesting to compare the lattice simulation results with the corresponding continuum calculations. 

\acknowledgments

We would like to thank Simon Catterall, Poul Damgaard, Nima Doroud, Joel Giedt, So Matsuura, Fumihiko Sugino, David Tong and Mithat \"Unsal for comments and discussions. This work was supported in part by the European Research Council under the European Union's Seventh Framework Programme (FP7/2007-2013), ERC grant agreement STG 279943, ``Strongly Coupled Systems".

\appendix

\section{Absence of Spectrum Doublers in the 5D Lattice MSYM}
\label{sec:absence-doublers}

In this appendix we show the absence of bosonic and fermionic doublers at the corners of the Brillouin zone of the five-dimensional lattice theory\footnote{A similar analysis for the case of four-dimensional $\cN=4$ lattice Yang-Mills has been carried out in Ref. \cite{Catterall:2011pd}.}. It is straightforward to carry out the analysis in momentum space. On a five-dimensional hypercubic lattice a generic field $\Phi(\bx)$ has the Fourier expansion
\beq
\Phi(\bx) = \frac{1}{(La)^5} \sum_{\bp} e^{i \bp \cdot \bx} \Phi_{\bp},
\label{F-transform}
\eeq
where $L$ is the size of the lattice and $a$ the lattice spacing. We have $\bx = a \sum_{b=1}^5 n_b \hatbmu_b$ denoting the position on the hypercubic lattice. The momenta $\bp$ lie on the dual lattice given by $\bp = \frac{2\pi}{La} \sum_{b=1}^5 m_b \hatbnu_b$. The dual basis vectors $\hatbnu_b, b = 1, \cdots, 5$ satisfy 
\beq
\hatbmu_a \cdot \hatbnu_b = \delta_{ab}.
\eeq

On an $L^5$ lattice both sets of lattice coordinates, $n_b$ and $m_b$, take integer values in the range $-L/2+1, \cdots, L/2$. We will assume periodic boundary conditions in all directions in the analysis. Eq. (\ref{F-transform}) implies that the fields are automatically invariant under translations by a lattice length in any direction and a field shifted by one of the basis vectors can be expressed in the following form\footnote{For simplicity we will adopt the convention that momentum sums $\sum_\bk$ automatically include the normalization factor of $1/(La)^5$.}
\beq
\Phi(\bx + \hatbmu_b) = \sum_{\bp} e^{ip_b} e^{i \bp \cdot \bx} \Phi_{\bp},
\eeq
where $p_b = \frac{2\pi}{L} m_b$. 

The bosonic action when expanded around unit links
\beq
\label{eq:U-expansion}
\cU_a = {\bf I} + \cA_a(\bn),~~~~\cUb_a = {\bf I} + \cAb_a(\bn),
\eeq
gives the following second-order term in Fourier space
\bea
S^{(2)}_{B} &\approx& \sum_{\bk,c,d,e} \Tr \Big(\cAb_c(\bk) \Big[\delta_{cd} f_e(\bk) f^*_e(\bk) - f^*_c(\bk) f_d(\bk) \Big] \cA_d(-\bk) \nn \\
&&\quad \quad \quad \quad + \phi_c(\bk)~\Big[ f^*_c(\bk) f_d(\bk) \Big]~\phi_d(-\bk) \Big),
\eea
where
\beq
f_e(\bk) \equiv (e^{i k_e} - 1).
\eeq

We need to gauge-fix the bosonic action before we derive the propagators. A natural gauge-fixing choice would be an obvious generalization of Lorentz gauge-fixing \cite{Marcus:1995mq}
\beq
G(\bn) = \sum_c \Big( \partial_c^{(-)} \cA_c(\bn) + \partial_c^{(-)} \cAb_c(\bn) \Big).
\label{g-fix}
\eeq

This gauge-fixing choice adds the following term to the bosonic action at quadratic order
\beq
S_{GF} = \frac{1}{4 \xi} \sum_{\bn} G^2(\bn) = \frac{1}{\xi} \sum_{\bn, c} \Tr (\partial^{(-)}_c A_c(\bn))^2,
\eeq
where $\partial^{(-)}_c f(\bn) = f(\bn) - f(\bn - \hatbmu_c)$ and $\xi$ the gauge fixing parameter. 

On using the relation \beq
\sum_{\bn} (\partial^{(+)}_c f(\bn)) g(\bn) = -\sum_{\bn} f(\bn) \partial^{(-)}_c g(\bn),
\eeq
the gauge-fixing term becomes
\beq
S_{GF} = -\frac{1}{\xi} \sum_{\bn,a,b} \Tr A_a(\bn) \partial^{(+)}_{a} \partial^{(-)}_b A_b(\bn).
\eeq

In momentum space it takes the form
\bea
S_{GF} &=&  \frac{1}{\xi} \sum_{\bk,a,b} \Tr A_a(\bk) \Big( f^*_a(\bk) f_b(\bk) \Big) A_b(-\bk).
\eea

Thus the gauge-fixed bosonic action to quadratic order is
\bea
S^{(2)}_B + S_{GF} &\approx& \sum_{\bk,a,b,c} \Tr \Big(A_a(\bk)~\Big[ \delta_{ab} f_c(\bk) f^*_c(\bk) - \Big(1 - \frac{1}{2\xi}\Big) f^*_a(\bk) f_b(\bk) \Big]~A_b(-\bk) \nn \\
&&\quad \quad \quad \quad + \phi_a(\bk)~\Big[ \delta_{ab} f_c(\bk) f^*_c(\bk) \Big]~\phi_b(-\bk) \Big).
\eea

The choice $\xi = 1/2$ makes the above expression diagonal
\bea
S^{(2)}_B &\approx& \sum_{\bk,a,b,c} \Tr \cAb_a(\bk)~[ \delta_{ab} f_c(\bk) f^*_c(\bk)]~\cA_b(-\bk) \nn \\
&=& \sum_{\bk,a,b} \Tr \Big[ \cAb_a(\bk) \delta_{ab} \Big( 4 \sum_c \sin^2 \Big(\frac{k_c}{2} \Big) \Big) \cA_b(-\bk)\Big].
\eea

Taking the trace the quadratic bosonic action can be written as
\beq
S^{(2)}_B \approx \sum_{\bk,a,b} \cAb^A_a(\bk) \Big(\widehat{\bk}^2 \delta_{ab} \delta_{AB}\Big) \cA^B_b(-\bk),
\eeq
where 
\beq
\widehat{\bk}^2 = 4 \sum_c \sin^2 \Big(\frac{k_c}{2}\Big).
\eeq

This leads to the following bosonic propagator in the theory 
\beq
\langle \cA_a^A(-\bk) \cAb_b^B(\bk)\rangle = \delta_{ab} \delta_{AB} \frac{1}{\widehat{\bk}^2}.
\eeq

Thus we see that $\widehat{\bk}^2 \neq 0$ at the edge of the five-dimensional Brillouin zone, $B = [-\pi, \pi]^5$.

Let us move on to computing the fermionic propagators. The fermionic part of the action is of the form
\bea
S_F &\approx& \sum_{\bn, a, b, c, d, e} \Tr \Big( \chi_{ab}(\bn) \cD^{(+)}_a\psi_b(\bn) + \eta(\bn) \cDb^{(-)}_a \psi_a(\bn) \nn \\
&&\quad \quad \quad \quad + \frac{1}{8} \epsilon_{abcde} \chi_{de}(\bn+\hatbmu_a+\hatbmu_b+\hatbmu_c) \cDb^{(+)}_c\chi_{ab}(\bn) \Big).
\eea

When expanded up to second order in the fields using Eq. (\ref{eq:U-expansion}), it becomes
\bea
S^{(2)}_F &\approx& \sum_{\bk,a,b,c,d,e} \Tr \chi_{ab}(\bk)\Big[-f^*_a(\bk) \delta_{bc} + f^*_b(\bk) \delta_{ac}\Big] \psi_c(-\bk) + \eta(\bk)f_c(\bk)\psi_c(-\bk) \nn \\
&&\quad \quad \quad \quad + \frac{1}{8} \epsilon_{abcde} \chi_{de}(\bk) e^{i(k_a+k_b)} f_c(\bk) \chi_{ab}(-\bk).
\eea

Upon restricting the sum and rescaling the field $2\chi_{ab} \rightarrow \chi_{ab}$ the fermionic action becomes
\bea
S^{(2)}_F &\approx& \sum_{\bk,a<b; c,d<e} \Tr \Big(\chi_{ab}(\bk) \Big[-f^*_a(\bk) \delta_{bc} + f^*_b(\bk) \delta_{ac} \Big] \psi_c(-\bk) + \eta(\bk) f_c(\bk) \psi_c(-\bk) \nn \\
&&\quad \quad \quad \quad + \frac{1}{8} \epsilon_{abcde} \chi_{de}(\bk) e^{i(k_a+k_b)} f_c(\bk) \chi_{ab}(-\bk) \Big).
\eea

We can write the above action in the form of a matrix product
\bea
S^{(2)}_F &\approx& \sum_{\bk} \left(
\Psi (\bk) \Psi(-\bk)
\right) \left( \qtr \right)
\left(
\begin{tabular}{cc}
0 & $M(\bk)$ \\
$-M^T(\bk)$  & 0  \\
\end{tabular}
\right)
\left(
\begin{tabular}{c}
$\Psi(\bk)$ \\
$\Psi(-\bk)$ 
\end{tabular}
\right)
\nn \\ &=&
\qtr \sum_{\bk} \Phi(\bk) \mathcal{M} \Phi(\bk)
\label{eq:mat-prod}
\eea
where $\Phi\equiv(\Psi(\bk), \Psi(-\bk))$ and $\Psi_i = (\eta, \psi_1, \cdots, \psi_5, \chi_{12}, \cdots, \chi_{15}, \cdots, \chi_{45})$ and $M(\bk)$ is given in block matrix form
\begin{small}
\beq
\left(
\eta \ \psi_{a} \ \chi_{de}
\right)(\bk)
\left(
\begin{tabular}{ccc}
0 & $f_b(\bk)$ & 0 \\
$-f^*_a(\bk)$  & 0 & $ f_g(\bk) \delta_{ha} - f_h(\bk) \delta_{ga}$ \\
0 & $-f^*_d(\bk) \delta_{eb} + f^*_e(\bk) \delta_{db}$ & 
$\epsilon_{ghcde} q_{gh} f_c(\bk)$
\end{tabular}
\right)
\left(
\begin{tabular}{c}
$\eta$ \\
$\psi_{b}$ \\
$\chi_{gh}$
\end{tabular}
\right)(-\bk).\nn
\eeq
\end{small}
where $q_{gh} = e^{i(k_g + k_h)}$. We also note that $M$ has the properties $M^T(\bk) = -M^*(\bk) = -M(-\bk)$. 

From above the fermionic propagator matrix has the form 
\beq
\langle \Psi_i^A(\bk) \Psi_j^B(-\bk) \rangle = 2 M^{-1}_{ij}(\bk) \delta_{AB}.
\eeq

Upon using Mathematica to compute the inverse of the fermionic propagator matrix we obtain
\beq
M^{-1}(\bk) = 
\frac{1}{\widehat{\bk}^2} \left(
\begin{tabular}{ccc}
0 & $f_b(\bk)$ & 0 \\
$-f^*_a(\bk)$  & 0 & $ f_g(\bk) \delta_{ha} - f_h(\bk) \delta_{ga}$ \\
0 & $-f^*_d(\bk) \delta_{eb} + f^*_e(\bk) \delta_{db}$ & 
$-\epsilon_{ghcde} q^*_{de} f^*_c(\bk)$
\end{tabular}
\right).
\eeq

In order to write down the propagators we need to undo the earlier rescaling of the $\chi$ field giving a factor of $\hf$ in the $\psi - \chi$ propagators and a factor of $\qtr$ in the $\chi - \chi$ propagators. Thus the fermionic propagators are 
\bea
\langle \eta^A(\bk) \psi_b^B(-\bk) \rangle &=& \delta_{AB} \frac{2}{\widehat{\bk}^2} (e^{ik_a} - 1), \\
\langle \psi_a^A(\bk) \chi_{bc}^B(-\bk) \rangle &=& \delta_{AB} \frac{1}{\widehat{\bk}^2} \left[(e^{ik_b} - 1) \delta_{ac} - (e^{ik_c} - 1) \delta_{ab}\right], \\
\langle \chi_{ab}^A(\bk) \chi_{de}^B(-\bk) \rangle &=& -\delta_{AB} \frac{1}{2\widehat{\bk}^2} \epsilon_{abcde} e^{-i(k_a + k_b)} (e^{-ik_c} - 1).
\eea

It is important to see that the fermionic propagators do not contain doublers at the edge of the Brillouin zone. 


\end{document}